\documentclass[ aps,prl,amsmath,amssymb,reprint,showpacs]{revtex4-1}

\usepackage{verbatim}
\usepackage{hyperref}
\usepackage{graphicx}
\usepackage{dcolumn}
\usepackage{bm}

\begin{document}

\title{Highly-anisotropic elements for acoustic pentamode  applications}%

\author{Christopher N. Layman}%
\email{christopher.layman.ctr@nrl.navy.mil}
\author{Christina J. Naify}%
\affiliation{National Research Council, Washington, DC 20001, USA }
\date{\today}
	
\author{Theodore P. Martin}
\author{David C. Calvo}
\author{Gregory J. Orris}%
\affiliation{U.S. Naval Research Laboratory, Washington, DC 20375, USA}
\date{\today}%

\begin{abstract}
Pentamode metamaterials are a class of acoustic metafluids that are characterized by a divergence free modified stress tensor.  Such materials have an unconventional anisotropic stiffness and isotropic mass density, which allow themselves to mimic other fluid domains.  Here we present a pentamode design formed by an oblique  honeycomb lattice and producing customizable anisotropic properties.  It is shown that anisotropy in the stiffness can exceed three orders of magnitude, and that it can be realistically tailored for transformation acoustic  applications.%
\end{abstract}

\maketitle

\textit{Introduction}.~Metamaterials comprise a relatively new area of research, which utilizes a wave-functionalized microstructure  to yield  unconventional effective material properties.  With its origin in microwave optics \citep{Veselago1968}, domains of study have since expanded to include near-infrared and visible optics \cite{Zhang4922,Dolling53}, radio-frequency electronics \cite{PhysRevLett.108.193902} and magnetics \cite{Kobljanskyj:2012fk}, with both linear and nonlinear \cite{Shadrivov:2008gd} phenomena being addressed.  
  
Many of the exotic phenomena observed in optical metamaterials, such as negative refraction \cite{Shelby:2001dk} and near-zero permittivity \cite{Edwards:2008kl},  have inspired analogous studies in acoustics.  Both acoustic \cite{Sukhovich:2008pi} and elastodynamic \cite{Lee:2011cs} negative refraction have been experimentally observed.  Narrowband broad-angle negative refraction was achieved using simultaneous negative mass density (dipole) and stiffness (monopole) \cite{Li:2004qa,Ding:2007rw,Wu:2011mi}, or by novel anisotropy \cite{Christensen:2012oz} utilizing perforated plates.

Conversely, coordinate transformation methods \cite{chan2010} have generated many new acoustic device concepts that do not require material negativity, such as lenses \cite{layman:163503}, beam splitters \cite{Bucay:2009jt}, black holes \cite{chang:054102} and scattering reducing cloaks \cite{Chen:2007ss,PhysRevB.83.224304,PhysRevLett.108.014301,Popa:2011ay}.  Such devices are defined as metafluids, which are effective materials with unconventional fluid-like properties whose particular bulk realization typically requires an anisotropic mass density, or in some limiting cases  a tensor elasticity for flexural waves in thin plates \cite{Farhat:2012fk}.   Experimental demonstration has been sparse, with most studies relying on a superlattice approach of alternating isotropic layers \cite{Torrent:2010yo}.  However, such an approach is difficult to realize and limited by the so-called mass catastrophe, which requires infinite mass density in the effective material profile.

\begin{figure}[b!]
\centering
\centerline{\includegraphics[keepaspectratio]{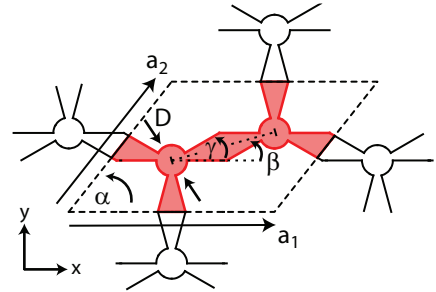}}
\caption{(Color online) Schematic of an anisotropic two-dimensional pentamode microstructure, with one unit cell highlighted.  Bulk effective properties were tuned with variation in cell geometry including joint diameter D, strut angle $\beta$ and inner-strut angle $\gamma$.  The unit cell angle $\alpha$ is obtained by $\alpha=\text{tan}^{-1}(1+\text{tan}(\beta))$, with $-15^\text{o}<\beta<30^\text{o}$.  $\mathbf{a}_1$ is held fixed throughout.}
\label{fig:unit_cell}
\end{figure}

An alternative approach is to generalize the conventional stress strain relationship to include pentamode elastic metamaterials.  Pentamode materials \cite{Milton:2002,Norris08092008,Norris:2009fk} are metafluids that support five easy infinitesimal strains (i.e. there is only one non-zero eigenvalue of the elasticity tensor which is of a pure pressure type), and satisfies the invariance of the governing equations by virtue of maintaining a harmonic transformation.  Pure pentamodes, in general, have an isotropic density and anisotropic stiffness with a negligible shear modulus.  Recently, isotropic pentamode materials have become experimental reality, and it was shown that the structure's effective bulk modulus exceeded the shear modulus by three orders of magnitude \cite{kadic:191901}.   However, it has yet to be reported that anisotropic pentamode metamaterials can be realistically implemented for specific applications, since an elastic solid with a zero shear modulus would have no stability and immediately flow away. 

 In this Letter, we show that an oblique honeycomb lattice can be utilized as a simple yet versatile building block for pentamode device construction, which exerts highly anisotropic control over sound waves.  The method presents a distinctly different approach to acoustic metamaterials, in that it does not require the difficult to achieve high value anisotropy in the effective mass density in addition to removing frequency bandwidth problems associated with inertial metafluids.  Potential applications include extraordinary scattering reduction and arbitrary wave manipulation, low loss acoustic delay lines \cite{yin:092905}, and phase controlled logic gates \cite{bringuier:1919}.

\begin{figure}[t!]
\centering
\centerline{\includegraphics[keepaspectratio]{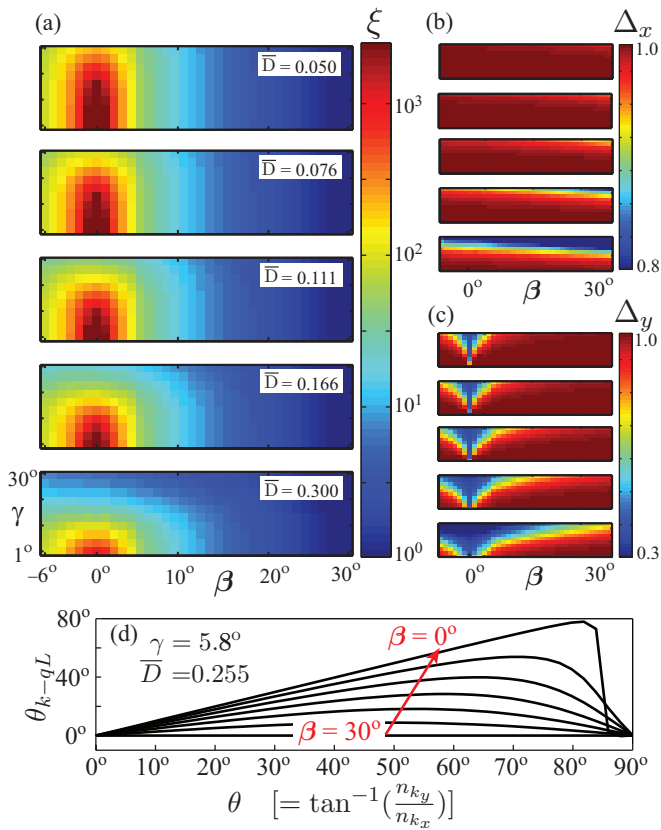}}
\caption{(Color online) Properties of an anisotropic pentamode metafluid. (a) Anisotropy factor $\xi$ as a function of unit cell geometric parameters, using the constituent properties of steel and lattice orientation as in Fig.~\ref{fig:unit_cell}.  Pentamode factor $\Delta_x$ in the stiff direction (b) and  $\Delta_y$ weak direction (c).  Note scale difference between (b) and (c), and that the y-axis in (b) and (c) is the same as in (a).  (d) Shows the behavior of quasi-modes as the design becomes more anisotropic (small $\beta$) at high propagation angles ($\theta$, which is measured from the horizontal axis).  The volume fraction for the range of parameters covered in (a-d) spans from approximately 1.5\% to 64\%.}
\label{fig:anisotropy}
\end{figure}

\textit{Anisotropy in pentamode metafluids}.~We consider elastic wave propagation in a microstructure having the general characteristics presented in Fig.~\ref{fig:unit_cell}.  For simplicity we present our results in a two-dimensional (2D) plain strain space, however, the analysis can straightforwardly be extended to three dimensions.   When the propagating wavelength is much larger than the lattice constants $\mathbf{a}_1$ and $\mathbf{a}_2$, the microstructure can be assigned bulk homogeneous effective properties.  As shown in Fig.~\ref{fig:unit_cell} the strut angle $\beta$ can be tuned to create isotropic ($\beta\sim30^\text{o}$), highly anisotropic ($\beta\sim0^\text{o}$) and re-entrant ($\beta<0^\text{o}$) materials.  Hence, based upon a targeted property profile, a gradient structure can then be designed having smoothly varying properties related to the local changing microstructure.  The condition that $\mathbf{a}_1$ remain constant throughout does not effect the generality of the results presented.

To demonstrate the properties of the structure, finite element analysis (COMSOL Multiphysics) was used to determine the structural eigenfrequenies of each geometric permutation, which has Bloch-Floquet periodic conditions prescribed on the unit cell boundaries.   This is analogous to computing the elastic wave band structure where the first two eigenfrequencies correspond to quasi-shear (qS) and quasi-longitudinal (qL) modes, respectively (see below for discussion on quasi-modes). These two eigenfrequencies were selected at small values of the wave number, near the origin where the dispersion curves are linear,  to obtain long wavelength  quasi-shear $v_{qS}$ and quasi-longitudinal $v_{qL}$ phase speeds.  An anisotropy factor $\xi$ is then defined as the ratio of effective bulk modulus in the principle directions,
\begin{equation}\label{eq:xi}
\xi = \frac{\text{K}_{x}}{\text{K}_{y} }= \frac{[v_{qL}^2-v_{qS}^2]_{x}}{[v_{qL}^2-v_{qS}^2]_{y}}
\end{equation}
Similarly, a pentamode factor $\Delta$, which assess the contribution of shear, is defined as 
\begin{equation}\label{eq:delta}
\Delta_{x,y} =  [\frac{v_{qL}^2- v_{qS}^2}{v_{qL}^2}]_{x,y}
\end{equation}
where the subscript represents the propagation direction.  Inversion methods \cite{Fokin:2007qc} have also been widely used to retrieve metamaterial effective properties, however such a method retrieves only a single modulus and is sensitive to termination conditions in the finite metamaterial slab used in the simulation.

In the above equations it is assumed that the effective dynamic mass density $\widetilde{\rho}$ is isotropic and equal to the  volume averaged mass density (and hence is removed from Eqs.~(\ref{eq:xi}-\ref{eq:delta})).  However, this assumption must be justified since  $\widetilde{\rho}$ can be anisotropic \cite{Popa:2009jo}.  Isotropy was verified by calculating the global reactant  force and acceleration on the unit cell for a given harmonic displacement in the principle directions.  It was concurrently determined that $\widetilde{\rho} $ was frequency independent within the long wavelength limit (see Supplemental Material).

Figure \ref{fig:anisotropy}(a) displays the calculated anisotropy factor, using the material properties of steel for the lattice and void for the interstitials, as a function of  the geometric parameters $\beta$, $\gamma$ and $\overline{\text{D}}$, where $\overline{\text{D}} = \text{D}/ |\mathbf{a}_1|$.  Exponentially increasing anisotropy is seen as $\beta$ approaches 0$^o$, with $\xi$ exceeding 10$^3$ for  small $\gamma$ and $\overline{\text{D}}$.  Figures 2(b)-(c) display the calculated pentamode factor, for the same geometric conditions as in Fig.~2(a).   When $\beta$ is small, $\Delta$ itself can be anisotropic; the contribution to shear modes is negligible in the x direction (Fig.~2(b)) and substantial in the y direction (Fig.~2(c)).  This occurs because as $\beta$ decreases the compression modulus ($\sim v_{qL}^2$) also decreases in the weak direction, yet the shear modulus ($\sim v_{qS}^2$) remains constant.  

In isotropic media, the polarization direction of the propagating wave is either parallel or normal to the wave vector direction.  In anisotropic media, however, this is in general not the case and waves propagate as quasi-longitudinal (qL) and quasi-shear (qS) modes.  The angle between the propagation direction $\mathbf{n}_k=\mathbf{k}/|\mathbf{k}|$ and the qL polarization direction $\mathbf{n}_{qL}$  is expressed as
\begin{equation}
 \theta_{k-qL} = \text{cos}^{-1}(\mathbf{n}_k \cdot \mathbf{n}_{qL}).
\end{equation}
Since qL and qS modes are orthogonal (i.e. $\mathbf{n}_{qS} \cdot \mathbf{n}_{qL}=0$) $ \theta_{k-qS} =  \theta_{k-qL}+\frac{\pi}{2}$, and for pure modes $ \theta_{k-qL}=0$. It is seen in Fig.~2(d)  that under the current design there is a very  strong dependence of $\theta_{k-qL}$ on $\beta$, and that for small $\beta$ there exist an extreme resistance of the wave to displace out of the stiff ($x$) direction.  In summary, there are two phenomena which have the potential to affect material performance from the pentamode ideal, and both are born out of the relationship between the stiff (stretch-dominated) and weak (bending-dominated) directions of the lattice in the extreme anisotropic cases when $\beta$ is small.    One (Fig.~\ref{fig:anisotropy}(c)) involves an increasing role of the shear mode in the weak ($y$) direction due to a decreasing compression modulus,  and the other (Fig.~\ref{fig:anisotropy}(d)) involves extreme deviation from pure propagating modes which tend to polarize along the stiff ($x$) direction for qL and weak ($y$) for qS.  

\begin{figure}[t!]
\centering
\centerline{\includegraphics[keepaspectratio]{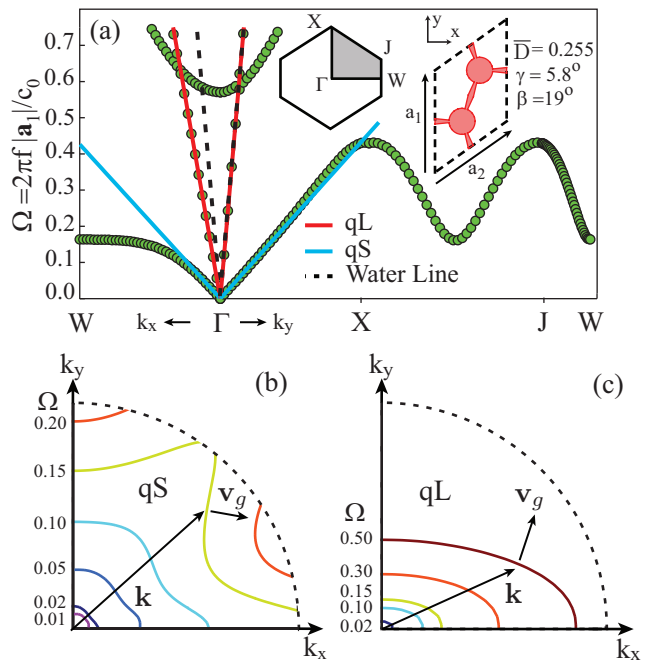}}
\caption{(Color online) (a) Normalized band structure along the principal directions of propagation for the unit cell used in the acoustic mirage.  Inset shows the unit cell and  first Brillouin zone with the irreducible part shaded in gray. (b)-(c) Equifrequency contours for qS and qL modes at selected frequencies (contours not calculated to zone boundaries).  The dotted curve depicts the wave vector $\mathbf{k}$  direction, and  anisotropic behavior is depicted by the difference between $\mathbf{k}$ and group velocity vector $\mathbf{v}_g$}
\label{fig:band_structure}
\end{figure}

\begin{figure*}[t!]
\centering
\centerline{\includegraphics[keepaspectratio]{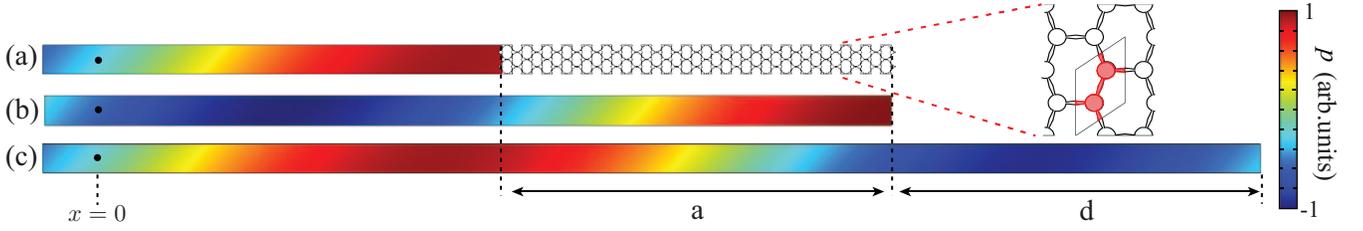}}
\caption{(Color online)  Acoustic mirage device for a mirage parameter of $a=d$, showing the real part of the scattered pressure field, using water as the background fluid.  Shown is the case for frequency $\Omega =0.1$ and an incident angle $\theta_i=43.4^o$, geometric parameters are those from Fig.~\ref{fig:band_structure}.  An observer at $x=0$ sees the same acoustic field (amplitude and phase) with the pentamode metafluid (a) as without (c).  For reference, a domain (b) is shown with the same length as the mirage. A plane wave travels from left to right, with imposed zero-reflection condition on left boundaries and zero displacement condition on right boundaries.  Top and bottom boundaries have Bloch-Floquet conditions.}
\label{fig:mirage}
\end{figure*}

\textit{Device applications}.~To demonstrate the efficiency of our method to devise acoustic metafluids, we employ it to design two different transformation acoustic devices. The first one is based on the concept of an acoustic mirage, where an observer hears an echo from a distant wall, whereas in reality the echo originates from a much closer boundary.  Norris \cite{Norris:2009fk} theoretically showed that a  2D mirage can be implemented using a pure pentamode metamaterial having orthotropic symmetry in the stiffness tensor.  This metafluid, of length $a$, mimics a larger fluid domain, of length $a+d$, and is both impedance matched and has equivalent acoustic travel times, such that to an external observer the two domains are indistinguishable (see Fig.~\ref{fig:mirage}).  The mirage  density and bulk modulus are expressed as \cite{Norris:2009fk}
\begin{equation}\label{eq:penta}
\widetilde{\rho} =\frac{a+d}{a}\rho_0, \quad\quad  K = K_0 \begin{bmatrix} \frac{a}{a+d} & 0 \\ 0 & \frac{a+d}{a}  \end{bmatrix}
\end{equation}
where $\rho_0$ and $K_0$ are the density and bulk modulus of the background fluid, in the present case water.  Equation (\ref{eq:penta}) describes a fluid with an orthotropically symmetric K, and  the mirage device can now be designed by  choosing a constituent material and adjusting the unit cell geometry to give the required properties.

Figure \ref{fig:band_structure}(a) shows the  band structure of a unit cell (see inset of Fig.~\ref{fig:band_structure}(a)) which approximately satisfies Eq.~(\ref{eq:penta}) for a mirage parameter of $a=d$, as a function of the normalized frequency $\Omega =  2\pi f |\mathbf{a}_1|/c_0$, where $f$ is the circular frequency and $c_0$ is a reference sound speed.   For this choice of geometric parameters, K$_{11} = 0.5052$K$_0$, K$_{22} = 2.0\text{K}_0$  and $\widetilde{\rho}=2.037 \rho_0$, as can be seen in the band structure.  With using silver as the constituent material,  $\xi=0.24$, $\Delta_x = 0.96$  and $\Delta_y=0.99$ (note that the unit cell of Fig.~\ref{fig:band_structure} is rotated 90$^o$ as compared to that in Fig.~\ref{fig:unit_cell}).  The device has a sufficient amount of shear for structural stability yet still retains pentamode characteristics.  Additionally, a wideband shear-wave directional band gap is noticed for propagation in the weak (x) direction.  Figure~\ref{fig:band_structure}(b-c) shows the equifrequency contours of the qS and qL modes at selected frequencies. At $\Omega < 0.01$ the qS modes are  approximately isotropic, with deviation from circularity of less than 5\%.  For higher $\Omega$ the qS modes have both significant increasing anisotropy and dispersion, as indicated by the relationship between group velocity and wave vectors inside the lattice.  The equifrequency contours for the qL mode, conversely,  maintains the required ellipsoidal anisotropy over a broad range of $\Omega$.

\begin{figure}[b!]
\centering
\centerline{\includegraphics[keepaspectratio]{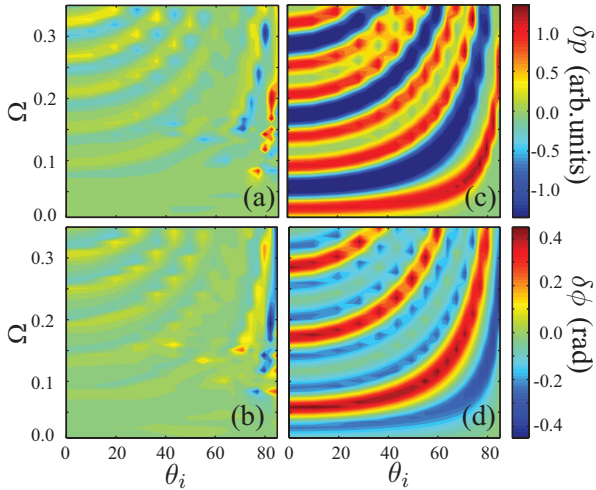}}
\caption{(Color online) Difference of scattered pressure amplitude ($\delta p$) and phase ($\delta \phi$) of the mirage (a) and (b), and reference domain (c) and (d), evaluated at a point location far from the water-metafluid interface, $x=0$. Plotted as a function of normalized frequency $\Omega$ and incident angle $\theta_i$.}
\label{fig:mirage_summary}
\end{figure}

Using a unit cell with the specific properties obtained above, an acoustic mirage device is constructed as depicted in Fig.~\ref{fig:mirage}.  Figures \ref{fig:mirage}(a) and \ref{fig:mirage}(c)  respectively compares the scattered acoustic fields for the mirage and the non-mirage case (the larger domain, termed the control), and they are seen to have excellent agreement.  A reference domain that is the same size as the mirage domain is also shown in Fig.~\ref{fig:mirage}(b).   Since the pentamode design is not based on narrow resonances, the mirage device will also operate over both a very wide frequency bandwidth and all angles of incidence.  Figure \ref{fig:mirage_summary}(a) and \ref{fig:mirage_summary}(b) summarizes the difference of the scattered pressure field amplitude $\delta p$ and phase $\delta \phi$ between the mirage and control, with excellent agreement seen.    This is compared to Figs.~\ref{fig:mirage_summary}(c) and \ref{fig:mirage_summary}(d), which depend strongly on the incident angle and frequency between the two domains.

\begin{figure}[b!]
\centering
\centerline{\includegraphics[keepaspectratio]{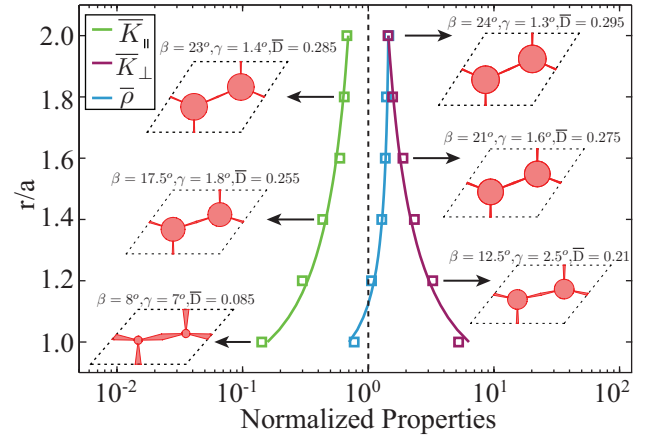}}
\caption{(Color online)  Designing material properties for transformation acoustics scattering reduction.  Solid lines denote ideal pentamode properties and symbols are properties achieved under the current design.  All values are normalized to water.}
\label{fig:cloak}
\end{figure}

The above outlined design method also has the flexibility to realize more complicated material profiles.  To demonstrate this, we use the method to realize the estimated material property values for a cylindrical pentamode scattering reduction layer for use in an aqueous environment.  A graded anisotropic honeycomb lattice is used to simulate the transformed space,  which reduces the actual size of the cylinder's radius $a$ to virtual radius $\delta$.   One possible mapping function (though not unique)  describing the transformation layer is \cite{gokhale:2932} 
\begin{equation}\label{eq:TA_function}
f(r)=(b^2-a^2)^{-1} \Big{[}(b^2-a\delta)r-(a-\delta)b^2 \frac{a}{r}\Big{]}
\end{equation}
where $r$ and $b$  are the radial coordinate and outer radius of the transformation layer, respectively.  Equation (\ref{eq:TA_function}) maintains both radial impedance and orthogonal wave speed matching with the external fluid.  Figure \ref{fig:cloak} compares the ideal pentamode material properties and values obtained from the composite metafluid for the case of $\delta=a/3$ and $b=2a$, using steel as the base material.  In 2D, the pure pentamode properties can be obtained through \cite{Norris08092008}
\begin{align}\label{eq:cylinder_cloak}
\overline{K}_\shortparallel(r)& = \frac{1}{f'(r)} \frac{f(r)}{r}\nonumber \\ \overline{K}_\perp(r)&=f'(r)\frac{r}{f(r)} \quad\quad a\leq r\leq b   \\   \overline{\rho}(r)&=f'(r) \frac{f(r)}{r} \nonumber
\end{align}
where the over line represent normalization to water values. Excellent approximation is demonstrated.  For all cases of the  designed  properties, pentamode behavior is achieved, $\Delta_x=\Delta_y\sim 1$.   At $r=a$, in Fig.~\ref{fig:cloak}, $\overline{K}_\shortparallel = 0.16$, $\overline{K}_\perp = 6.33$, and $\overline{\rho} = 0.70$.  An inertial metafluid version of the transformation layer using Eq.~(\ref{eq:TA_function}) produces (at $r=a$)  $\overline{K} = 1.42$, $\overline{\rho}_\perp = 0.157$, and $\overline{\rho}_\shortparallel = 6.33$.  Using a superlattice approach to realize these inertial values requires alternating isotropic layers with properties $\overline{\rho}_1=12.58$, $\overline{K}_1=6.29$ and $\overline{\rho}_2=0.079$, $\overline{K}_2=0.80$.  Realizing such a composite material is prohibitively  difficult.

Finally, it is noted that the approximation of the ideal properties in Eq.~(\ref{eq:cylinder_cloak}) are for a cylindrical coordinate system, whereas the unit cell analysis is done in a rectilinear system.  However, it is the point of this work to demonstrate the feasibility of the design to approximate unique property values required in pentamode  applications, the design of a gradient curved structure based upon an anisotropic honeycomb lattice is beyond the scope of the current work, and will be presented in a future publication.  Furthermore, since the current method is fully scalable, as the size of the unit cell is decreased the local difference  between the curvilinear and rectilinear coordinate systems will be reduced.  

\textit{Conclusions}.~A versatile method to design and characterize highly anisotropic pentamode elements for transformation acoustics was presented.  The design method is based upon customizing the geometric parameters of an oblique honeycomb lattice to target particular transformation acoustics applications.  In the low frequency limit, the metallic structures act as  acoustic metafluids with pentamode properties, and have negligible shear modulus relative to its anisotropic fluid-like stiffness.  The simplicity of the design (easy to fabricate) coupled with the effectiveness of performance shows promise to realize concepts hitherto constrained by inertial-based metamaterial constructions.  

This work was supported by the Office of Naval Research.


%
%

%

\pagebreak

\section{Supplemental Material}
\begin{figure*}[b!]
\centering
\centerline{\includegraphics[keepaspectratio]{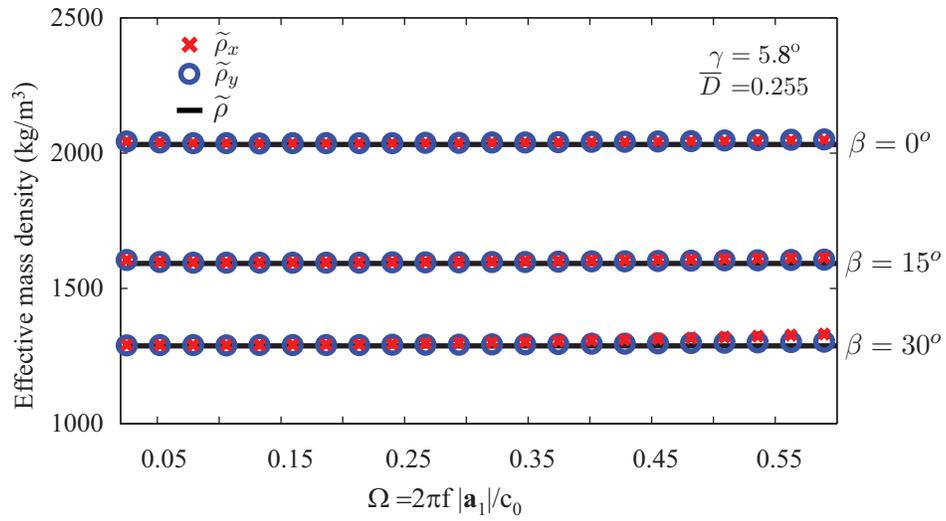}}
\caption{(Color online) Comparison between volume averaged mass density $\widetilde{\rho}$ and anisotropic mass densities $\widetilde{\rho}_x$ and $\widetilde{\rho}_y$.  Equivalence is shown within the long wavelength approximation.}
\label{fig:SMrho}
\end{figure*}

 To verify isotropic mass density, the global reactant force $F_i$ and acceleration $a_i$ is obtained by averaging the local surface traction $T_j = n_i  \sigma_{ij}$ and acceleration $\ddot{u_{i}}$ on the unit cell's external boundary $\Sigma$, where $n_i$ and $\sigma_{ij}$ are the local surface unit normal vector and stress tensor, respectively.   These can be expressed as, $F_j = \frac{1}{A}\int_{\Sigma} n_i \sigma_{ij}dr \ \text{and} \ a_i = \frac{1}{\ell}\int_{\Sigma} \ddot{u_i}\ dr$, where $A, \ell$ and $dr$ are the area of the unit cell, total path length of integration and differential length of integration along $\Sigma$, respectively.  The effective dynamic anisotropic mass density is therefore obtained simply as $\widetilde{\rho_i} ={F_i}/{a_i}$.  Figure \ref{fig:SMrho} compares $\widetilde{\rho}$ and  $\widetilde{\rho_i}$ ($i=x,y$), using steel as the base material,  and it is seen that they are approximately equivalent.     $\widetilde{\rho}_i$ is also frequency independent within the long wavelength limit.  

\end{document}